\documentclass[12pt]{article}
\usepackage{amsmath,epsfig,amssymb,bbm,amssymb}

\newcommand{\jtheta}[1]{\vartheta \begin{bmatrix} #1 \end{bmatrix}}

\begin{document}

\title{\vbox{
\baselineskip 14pt
\hfill \hbox{\normalsize WU-HEP-09-05} \\
\hfill \hbox{\normalsize KUNS-2249}\\
\hfill \hbox{\normalsize YITP-09-122} } \vskip 2cm
\bf 
Flavor structure from  
magnetic fluxes and non-Abelian Wilson lines \vskip 0.5cm
}
\author{
Hiroyuki~Abe$^{1,}$\footnote{email:
 abe@waseda.jp}, \
Kang-Sin~Choi$^{2,}$\footnote{email:
  kschoi@gauge.scphys.kyoto-u.ac.jp}, \
Tatsuo~Kobayashi$^{2,}$\footnote{
email: kobayash@gauge.scphys.kyoto-u.ac.jp} \ \\ and \
Hiroshi~Ohki$^{2,3,}$\footnote{email: ohki@scphys.kyoto-u.ac.jp
}\\*[20pt]
$^1${\it \normalsize
Department of Physics, Waseda University, Tokyo 169-8555, Japan} \\
$^2${\it \normalsize
Department of Physics, Kyoto University,
Kyoto 606-8502, Japan} \\
$^3${\it \normalsize 
Yukawa Institute for Theoretical Physics, Kyoto University, 
Kyoto 606-8502, Japan}
}

\date{}

\maketitle
\thispagestyle{empty}

\begin{abstract}
We study the flavor structure of 
4D effective theories, which are derived from 
extra dimensional theories with 
magnetic fluxes and non-Abelian Wilson lines.
We study zero-mode wavefunctions and compute 
Yukawa couplings as well as four-point couplings.
In our models, we also discuss non-Abelian discrete flavor symmetries 
such as $D_4$, $\Delta(27)$ and $\Delta(54)$.
\end{abstract}

\newpage

\section{Introduction}

Recently, extra dimensional field theory, 
in particular string-derived one, plays an important 
role in particle physics and cosmology.
When we start with extra dimensional field theory, 
it is one of most important issues how to 
realize a chiral spectrum in four dimensional (4D) effective 
field theory.
The magnetic flux background is one of interesting ways to 
realize a 4D chiral theory.
Indeed, several field-theoretical models and string 
models, i.e. magnetized D-brane models, have been studied~\cite{Manton:1981es,
Witten:1984dg,Bachas:1995ik,Berkooz:1996km,Blumenhagen:2000wh,
Angelantonj:2000hi,Cremades:2004wa,Troost:1999xn,Choi:2009pv}.
Furthermore, magnetized D-brane models are T-duals of 
intersecting D-brane models.
Within the latter type of model building, 
a number of interesting models have been
constructed~\cite{Berkooz:1996km,
Blumenhagen:2000wh,Angelantonj:2000hi,
Aldazabal:2000dg,Blumenhagen:2000ea,Cvetic:2001tj}.\footnote{
See for a review \cite{Blumenhagen:2005mu} and references therein.}.

Wavefunction profiles of zero-modes are quasi-localized on the torus with 
magnetic flux background.
The number of zero-modes is determined by the size of 
magnetic flux.
Since we know zero-mode profile explicitly, 
we can compute concretely 3-point and higher order couplings of 
4D effective field theory by overlap integrals of 
zero-mode profiles~\cite{
Cremades:2004wa,DiVecchia:2008tm,Abe:2009dr,Antoniadis:2009bg,Camara:2009uv}.
That is an important aspect of magnetized extra dimensional models.
Moreover, such a 4D effective field theory can have Abelian 
and non-Abelian discrete flavor symmetries, 
which are originated from localization behavior of zero-modes 
in extra dimensions~\cite{Abe:2009vi}.\footnote{
Similar non-Abelian discrete flavor symmetries have been 
derived within the framework of heterotic orbifold models~\cite{
Kobayashi:2004ya,Kobayashi:2006wq,Ko:2007dz}.
Analysis on their anomalies are also important~\cite{Araki:2008ek}.}

In addition to magnetic fluxes, we can introduce constant gauge 
backgrounds and non-trivial twisted boundary conditions as 
well as orbifold boundary conditions~\cite{
Cremades:2004wa,Abe:2008fi,Abe:2008sx,Abe:2009uz}.\footnote{
Geometrical backgrounds other than tori and toroidal orbifolds have
also been studied~\cite{Conlon:2008qi,Marchesano:2008rg,Camara:2009xy}.}
That leads richer structure in model building such as 
zero-mode spectra and zero-mode profiles.

Non-Abelian Wilson lines, i.e. the so-called toron backgrounds~\cite{toron},  
are also interesting backgrounds~\cite{
Guralnik:1997th,Cremades:2004wa,Alfaro:2006is,vonGersdorff:2007uz,Faedo:2009zz}.
They can break gauge groups with reducing their ranks.
For a certain case with magnetic fluxes and non-Abelian Wilson lines, 
zero-mode profiles have been given~\cite{Cremades:2004wa}.
Our purpose of this paper is to study more about models 
with magnetic fluxes and non-Abelian Wilson lines.
We analyze zero-mode profiles in generic case and compute 
3-point couplings.
Furthermore, we study flavor symmetries.
We also study the orbifold compactification.

This paper is organized as follows.
In section 2, we give a brief review on the extra dimensional 
models with magnetic fluxes and non-Abelian Wilson lines.
In section 3, we study zero-mode wavefunctions on 
the torus compactification with magnetic fluxes and 
non-Abelian Wilson lines.
We compute Yukawa couplings in section 4 and 
study flavor symmetries of our models in section 5.
In section 6, we also study the orbifold compactification.
Section 7 is devoted to conclusion and discussion.
In Appendix A we show useful calculations, which are relevant to 
Yukawa couplings and in Appendix B we compute 
four-point couplings as an example of higher order couplings.

\section{Non-Abelian Wilson lines} 

\subsection{Higher dimensional super Yang-Mills theory}

Our starting point is $N=1$ $U(N)$ super Yang-Mills theory in 
$D=4+2n$ dimensions with $n=1$ or 3.
Its Lagrangian is written by 
\begin{equation}
\label{eq:SYM-L}
{\cal L} = - \frac{1}{4g^2}{\rm Tr}\left( F^{MN}F_{MN}  \right) 
+\frac{i}{2g^2}{\rm Tr}\left(  \bar \lambda \Gamma^M D_M \lambda \right),
\end{equation}
where $M,N=0,\cdots, (D-1)$.
Here, $\lambda$ denotes gaugino fields, $\Gamma^M$ is the 
gamma matrix for $D$ dimensions and 
the covariant derivative $D_M$ is given as 
\begin{equation}
D_M\lambda = \partial_M \lambda - i [A_M, \lambda],
\end{equation}
where $A_M$ is the vector field.

Here, we consider the torus $(T^2)^n$ as 
$2n$-dimensional extra dimensions and denote their 
coordinates by $y_m$ $(m=4,\cdots,2n+3)$.
We use the orthogonal coordinates and choose the 
torus metric such that 
$y_m$ is identified by $y_m+1$ on the torus.
The gaugino fields $\lambda$ and the vector fields $A_m$ 
corresponding to the compact directions are decomposed as 
\begin{eqnarray}
\label{eq:gaugino-decomp}
\lambda(x,y) &=& \sum_n \chi_n(x) \otimes \psi_n(y), \\
\label{eq:vector-decomp}
A_m(x,y) &=& \sum_n \varphi_{n,m}(x) \otimes \phi_{n,m}(y),
\end{eqnarray}
where $x$ denotes the coordinates of four-dimensional 
uncompact space $R^{3,1}$.
Here, we are interested only in zero-modes, $\psi_0(y)$ and 
$\phi_{0,m}(y)$.
Thus, we omit the mode index corresponding to $n=0$ 
and write them as $\psi (y)$ and $\phi_{m}(y)$.

\subsection{Non-Abelian Wilson lines}

Here, we consider $T^2$ of $(T^2)^n$, whose coordinates are 
denoted as $(y_4,y_5)$.
As a $U(N)$ gauge background, we introduce the following form 
of (Abelian) magnetic flux,
\begin{equation}\label{eq:F45-UN}
F_{45} = 2 \pi \left(
\begin{array}{cc}
f_a {\bf 1}_{N_a} & 0 \\
0 & 0
\end{array}\right),
\end{equation}
where ${\bf 1}_{N_a }$ denotes $(N_a \times N_a)$ 
identity matrix.
For example, we use the following gauge, 
\begin{equation}\label{eq:gauge}
A_4 = -F_{45}y_5, \qquad A_5 = 0,
\end{equation}
for the $U(N_a)$ part.
Then, their boundary conditions can be written as 
\begin{eqnarray}
\label{eq:BC-gauge}
A_m(y_4+1,y_5)&=&A_m(y_4,y_5)+
\begin{pmatrix}
\partial_m \chi_4{\bf 1}_{N_a} & 0 \\
0 & 0
\end{pmatrix}
, \qquad   
\chi_4 = 0, \nonumber \\
A_m(y_4,y_5+1)&=&A_m(y_4,y_5)+
\begin{pmatrix}
\partial_m \chi_5 {\bf 1}_{N_a}& 0 \\
0 & 0 \\
\end{pmatrix}, \qquad 
\chi_5 = - 2 \pi f_a y_4.
\end{eqnarray}

This background breaks the gauge group $U(N)$ to 
$U(N_a) \times U(N-N_a)$.
The zero mode $\psi(y)$ corresponding to the gaugino is also 
decomposed as 
\begin{equation}\label{eq:psi-ABCD} 
\psi = 
\begin{pmatrix}
A & B \\
C & D 
\end{pmatrix},
\end{equation}
depending on their $U(N_a) \times U(N-N_a)$ charges.
That is, $A$ and $D$ correspond to the gaugino fields of 
unbroken symmetries, $U(N_a)$ and $U(N-N_a)$, respectively, 
while $B$ and $C$ correspond to bi-fundamental representations, 
$(N_a,\overline{N-N_a})$ and $(\overline{N_a},{N-N_a})$, respectively.

Since only the $U(1)$ part of $U(N_a)$  has the non-trivial background, 
its charge $q$ is relevant, that is, $A, B, C$ and $D$ have 
charges $q=0,1,-1$ and $0$, respectively.
For example, the zero-mode of $B$ elements satisfies the following equation,
\begin{equation}\label{eq:Dirac-T2}
\tilde \Gamma^m(\partial_m -iA_m)B(y) = 0,
\end{equation}
for $m=4,5$, where $A_m$ denotes the $U(N_a)$ gauge background 
(\ref{eq:gauge}).
Also, the zero-mode of $C$ elements  satisfies 
$\tilde \Gamma^m(\partial_m + iA_m)C(y) = 0$, 
while the zero-modes of $A$ and $D$ elements satisfy 
$\tilde \Gamma^m \partial_m A(y) = 0$ 
and $\tilde \Gamma^m \partial_m D(y) = 0$.
Here, $\tilde \Gamma^m$ corresponds to 
the gamma matrix for the two-dimensional torus $T^2$, e.g.
\begin{equation}
\tilde \Gamma^4 = \left(
\begin{array}{cc}
0 & 1 \\
1 & 0 
\end{array}
\right), \qquad 
\tilde \Gamma^5 = \left(
\begin{array}{cc}
0 & -i \\
i & 0 
\end{array}
\right),
\end{equation}
and $\psi(y)$ is the two component spinor,
\begin{equation}\label{eq:two-spinor}
\psi = \left(
\begin{array}{c}
\psi_+ \\ \psi_-
\end{array}
\right),
\end{equation}
that is, $A, B, C$ and $D$ also have two components, 
$A_\pm, B_\pm, C_\pm$ and $D_\pm$.

In particular, we are interested in matter fields.
Thus, let us concentrate on $B$ and $C$ fields. 
Because of (\ref{eq:BC-gauge}), the spinor field, e.g. $B$, satisfies 
the following boundary conditions,
\begin{eqnarray}
\label{eq:bc-1}
B(y_4+1,y_5) &=& e^{i\chi_4}B(y_4,y_5), \\
B(y_4,y_5+1) &=& e^{i\chi_5}B(y_4,y_5) .
\label{eq:bc-2}
\end{eqnarray}
Here, we write these boundary conditions as 
\begin{eqnarray}
B(y_4+1,y_5) &=& \Omega_4(y_4,y_5) B(y_4,y_5) , \\
B(y_4,y_5+1) &=& \Omega_5(y_4,y_5) B(y_4,y_5) .
\end{eqnarray}
The above case corresponds to  $\Omega_4(y_4,y_5)=e^{i\chi_4}$ and 
$\Omega_5(y_4,y_5)=e^{i\chi_5}$.
Then, the consistency for the contractible loop, i.e. 
$(y_4,y_5) \to (y_4+1,y_5) \to (y_4+1,y_5+1) \to (y_4,y_5+1) \to (y_4,y_5)$ 
 requires 
\begin{eqnarray}\label{eq:loop}
\left( \Omega_5^{-1}(y_4,y_5+1)  \Omega_4^{-1}(y_4+1,y_5+1) \Omega_5(y_4+1,y_5) 
\Omega_4(y_4,y_5) \right) B(y_4,y_5) = B(y_4,y_5).
\end{eqnarray}
The left hand side reduces to $e^{-2\pi i f_a}\psi(y_4,y_5)$
in the above background.
This  condition for $B$ leads to 
the quantization condition of the magnetic flux $f_a$. 
That is, the magnetic flux $f_a$ should be quantized such
that $f_a=$ integer.
The consistency condition for the $C$ fields also leads 
to the same quantization condition, i.e.   $f_a=$ integer.

When we introduce a non-trivial background for 
the $SU(N_a)$ part of $U(N_a)$, the situation changes.
That modifies the boundary conditions on, for example,  
$B$,
\begin{eqnarray}
B(y_4+1,y_5) &=& \Omega_4(y_4,y_5) B(y_4,y_5) = 
e^{i\chi_4} \omega_4 B(y_4,y_5), \\
B(y_4,y_5+1) &=& \Omega_5(y_4,y_5) B(y_4,y_5) = 
e^{i\chi_5} \omega_5 B(y_4,y_5),
\end{eqnarray}
where $\omega_m$ are constant elements of $SU(N_a)$.
Then, the consistency condition (\ref{eq:loop}) reduces to 
\begin{eqnarray}\label{eq:loop-na}
\omega_5^{-1} \omega_4^{-1}  \omega_5 \omega_4  e^{-2\pi i f_a} = 
{\bf 1}_{N_a }.
\end{eqnarray}
If $\omega_4$ and $\omega_5$ commute each other, 
that would require again  $e^{-2\pi i f_a} = 1$.
Thus, it is interesting that $\omega_4$ and $\omega_5$ do not 
commute each other, that is, non-Abelian Wilson lines.
In particular, we consider the case that 
$\omega_5^{-1} \omega_4^{-1}  \omega_5 \omega_4$ corresponds 
to the center of $SU(N_a)$, that is, 
\begin{eqnarray}\label{eq:twist-alg}
\omega_5^{-1} \omega_4^{-1}  \omega_5 \omega_4   = e^{ 2\pi i M_a/N_a}
{\bf 1}_{N_a },
\end{eqnarray}
where $M_a$ is an integer.
In this case, the consistency condition (\ref{eq:loop-na}) 
requires that the magnetic flux should satisfy $f_a= M_a/N_a$ 
(mod 1).

We denote $P_a = {\rm g.c.d.}(M_a,N_a)$, $m_a=M_a/P_a$ and 
$n_a=N_a/P_a$.\footnote{
Here, ${\rm g.c.d.}$ denotes the greatest common divisor.}
A solution of Eq.~(\ref{eq:twist-alg}) is given as 
\begin{eqnarray}\label{eq:nA-WL-1}
\omega_4 = \hat P_a, \qquad \omega_5 = \hat Q^{-m_a}_a,
\end{eqnarray}
where
\begin{eqnarray}\label{eq:nA-WL-2}
\hat P_a= 
\begin{pmatrix}
0 & {\bf 1}_{P_a } & 0 & 0 \\ 
0 & 0 & {\bf 1}_{P_a } & 0 \\
\cdots \\
{\bf 1}_{P_a } & 0 & 0 & 0 
\end{pmatrix}, \quad
\hat Q_a= \rho^{(n_a-1)/2}
\begin{pmatrix}
{\bf 1}_{P_a } & 0 & 0 & 0 \\ 
0 & \rho {\bf 1}_{P_a } & 0 & 0 \\
\cdots \\
0 & 0 & 0 & \rho^{n_a-1} {\bf 1}_{P_a }\\
\end{pmatrix},
\end{eqnarray}
with $\rho \equiv e^{{2\pi}i/{n_a}}$.

These non-Abelian Wilson lines break the gauge group $U(N_a)$ further.
The following condition on the $U(N_a)$ gauge field, 
\begin{eqnarray}
A_\mu = w_4 A_\mu \omega_4^{-1} = w_5 A_\mu \omega_5^{-1},
\end{eqnarray}
is required.
Then, the gauge group $U(N_a)$ breaks to $U(P_a)$.

\section{Matter fields}

Here, we consider the following form of $U(N)$ magnetic fluxes,
\begin{equation}\label{eq:F45-UN}
F_{45} = 2 \pi \left(
\begin{array}{ccc}
f_1 {\bf 1}_{N_1} & & 0 \\
 & \ddots & \\
0 & & f_n {\bf 1}_{N_n }
\end{array}\right).
\end{equation}
This form of magnetic fluxes breaks $U(N)$ to 
$\prod_i U(N_i)$ for $f_i=$ integer.
Furthermore, the gauge group is broken to 
$\prod_i U(P_i)$ when we choose $f_i=M_i/N_i$ with 
$P_i = {\rm g.c.d.}(M_i,N_i)$ and non-Abelian Wilson lines such that 
they satisfy the consistency condition like Eq.~(\ref{eq:loop}).

Now, let us focus on the $(N_a+N_b)\times (N_a+N_b)$ block 
in $U(N)$, which has the magnetic flux,
\begin{equation}\label{eq:mg-flux} 
F = 2 \pi
\begin{pmatrix}
\frac{m_a}{n_a} {\bf 1}_{N_a} &  \\
& \frac{m_b}{n_b} {\bf 1}_{N_b} \\
\end{pmatrix} .
\end{equation}
We use the same gauge as Eq.~(\ref{eq:gauge}), i.e.
\begin{eqnarray}
A_4= - 2\pi
\begin{pmatrix}
\frac{ m_a}{n_a} {\bf 1}_{N_a} & \\ 
 & \frac{ m_b}{n_b} {\bf 1}_{N_b} \\ 
\end{pmatrix}y_5,
 \quad \quad 
A_5= 0. 
\end{eqnarray}
Similarly to Eq.~(\ref{eq:BC-gauge}), we denote 
their boundary conditions as 
\begin{eqnarray}
A_m(y_4+1,y_5)&=&A_m(y_4,y_5)+
\begin{pmatrix}
\partial_m \chi^a_4 {\bf 1}_{N_a}& 0 \\
0& \partial_m \chi^b_4 {\bf 1}_{N_b}
\end{pmatrix},  \nonumber \\
A_m(y_4,y_5+1)&=&A_m(y_4,y_5)+
\begin{pmatrix}
\partial_m \chi^a_5 {\bf 1}_{N_a}& 0 \\
0& \partial_m \chi^b_5 {\bf 1}_{N_b}
\end{pmatrix},
\end{eqnarray}
where 

\begin{eqnarray}
\chi_4^a=0, \ \ \ 
\chi_5^a=-2\pi\frac{m_a}{n_a}y_4, \ \ \   
\chi_4^b=0, \ \ \ 
\chi_5^b=-2\pi \frac{m_b}{n_b}y_4 .
\end{eqnarray}

We decompose the gaugino fields of this block 
in a way similar to Eq.~(\ref{eq:psi-ABCD}).
That is, $A$ and $D$ correspond to adjoint matter fields 
of $U(N_a)$ and $U(N_b)$, respectively, while 
$B$ and $C$ correspond to bi-fundamental representations, 
$(N_a,\overline{N_b})$ and $(\overline{N_a},N_b)$, respectively.
Among them, we concentrate on the field $B$, 
which satisfies the boundary conditions,
\begin{eqnarray}\label{eq:B-BC}
B(y_4+1,y_5) &=&
\Omega_4^a B(y_4,y_5)  (\Omega_4^b)^\dagger =
e^{i(\chi_4^a-\chi_4^b)} \omega_4^a B(y_4,y_5) (\omega^b_4)^\dagger, 
\nonumber \\
 B(y_4,y_5+1) &=&
\Omega_5^a B(y_4,y_5)  (\Omega_5^b)^\dagger =
e^{i(\chi_5^a-\chi_5^b)} \omega_5^a B(y_4,y_5) (\omega^b_5)^\dagger.
\end{eqnarray}
Here, $\omega_{4,5}^{a,b}$ are non-Abelian Wilson lines, 
which are given as Eqs.~(\ref{eq:nA-WL-1}) and (\ref{eq:nA-WL-2}).
Then, the gauge symmetries are broken to $U(P_a)$ and $U(P_b)$.
We study zero-mode profiles of $B$ fields in what follows.

\subsection{Integer magnetic fluxes}
\label{sec:integer-mf}

Before considering the models with fractional magnetic fluxes 
and non-Abelian Wilson lines, it would be convenient to 
review briefly the models with integer magnetic fluxes  
and  no Wilson lines, that is, 
\begin{eqnarray}
n_a=n_b=1, \qquad w_4^{a,b}=w_5^{a,b}= {\bf 1}.
\end{eqnarray}
Then, the boundary condition \eqref{eq:B-BC} reduces to 
\begin{eqnarray}\label{eq:B-BC-0}
B_{pq}(y_4+1,y_5) &=& B_{p,q}(y_4,y_5), \nonumber \\ 
B_{pq}(y_4,y_5+1) &=& e^{-2\pi imy_4} B_{p,q}(y_4,y_5),
\end{eqnarray}
where $m=m_a-m_b$.
Suppose that $m>0$.
Then, the $B_+$ component for each $B_{p,q}$ element has 
$m$ independent solutions for the zero-mode Dirac equation 
(\ref{eq:Dirac-T2}) with above boundary condition \eqref{eq:B-BC-0}.
These solutions are given by 
\begin{eqnarray}
\Theta^{j}(y_4,y_5)
&=&
\sum_l e^{- m\pi(l+\frac{j}{m})^2 + 2\pi i
m(l+\frac{j}{m})y_4-{\pi m}y^2_5
-2\pi m(l+\frac{j}{m})y_5  } \nonumber  \\
&=&
e^{-{\pi m}y_5^2} \ 
\jtheta{\frac{j}{m} \\ 0}(mz,m\tau) ,
\end{eqnarray}
where $z=y_4+iy_5$, $j = 0, 1, \cdots, m-1$ and $\tau=i$.
Here, $\jtheta{\frac{j}{m} \\ 0}(mz,m\tau)$ denotes the 
Jacobi theta function.
On the other hand, the $B_-$ component has no normalizable zero-modes.
Similarly, the $C_-$ fields have the same solutions as $B_+$,
but the $C_+$ has no normalizable zero-modes.

When $m <0$, the $B_-$ and $C_+$ fields have the $|m|$ independent solutions 
with the same wavefunctions as above except replacing $m$ by $|m|$.
However, the $B_+$ and $C_-$ fields have no   normalizable zero-modes.

\subsection{Fractional magnetic fluxes}

Here, we study zero-mode profiles in the models with 
fractional magnetic fluxes and non-Abelian Wilson lines.

\subsubsection{$n_a = n_b$}

First, let us study the magnetic flux (\ref{eq:mg-flux}) 
for $n= n_a = n_b$.
In this case, the non-Abelian Wilson lines break
the gauge group $U(N_a)\times U(N_b)$ to 
$U(P_a) \times U(P_b)$, where $P_a=N_a/n$ and $P_b=N_b/n$.
Following this breaking pattern, we decompose the fields $B$ as 
\begin{eqnarray}
B= 
\begin{pmatrix}
B_{00} & B_{01} & \cdots & \\
B_{10} & B_{11} & \cdots & \\
\cdots \\
B_{n-1,0} & B_{n-1,1} & \cdots & B_{n-1,n-1} 

\end{pmatrix}.
\end{eqnarray}
Each of $B_{pq}$ components is  
$(P_a \times P_b)$ matrix-valued fields, 
which correspond to bi-fundamental $(P_a,\bar P_b)$
fields under $U(P_a)\times U(P_b)$.
The boundary condition (\ref{eq:B-BC}) due to 
the non-Abelian Wilson lines is written as
\begin{eqnarray}\label{eq:B-BC-1}
B_{pq}(y_4+1,y_5) &=& B_{p+1,q+1}(y_4,y_5), \nonumber \\ 
B_{pq}(y_4,y_5+1) &=& \rho^{-(m_ap-m_bq)}e^{-\frac{2\pi im}{n}y_4} 
B_{p,q}(y_4,y_5),
\end{eqnarray}
where $m$ is used as $m=m_a-m_b$, 
and $B_{p+n,q} = B_{p,q+n} = B_{p.q}$.
That leads to the boundary condition,
\begin{eqnarray}\label{eq:B-BC-n}
B_{pq}(y_4+n,y_5) &=& B_{pq}(y_4,y_5),  \nonumber \\ 
B_{pq}(y_4,y_5+n) &=& e^{-2\pi im y_4} B_{pq}(y_4,y_5).
\end{eqnarray}
Suppose that $mn>0$.
Then, similarly to section \ref{sec:integer-mf}, 
the $B_+$ component for $B_{p,q}$ has 
$nm$ independent solutions for the zero-mode 
Dirac equation (\ref{eq:Dirac-T2}) with the above 
condition (\ref{eq:B-BC-n}).
These solutions are given by 
\begin{eqnarray}\label{eq:solution-1}
\Theta^{j}(y_4,y_5)
&=&
\sum_l e^{- nm\pi(l+\frac{j}{nm})^2 + 2\pi i
m(l+\frac{j}{nm})y_4-\frac{\pi m}{n}y^2_5
-2\pi m(l+\frac{j}{nm})y_5  } \nonumber  \\
&=&
e^{-\frac{\pi m}{n}y_5^2} \ 
\jtheta{\frac{j}{nm} \\ 0}(mz,nm\tau) ,
\end{eqnarray}
where $j= 0, 1, \cdots, nm-1$ and $\tau=i$.
On the other hand, the $B_-$ component has no 
normalizable zero-modes.
One finds that these solutions satisfy 
the boundary conditions,
\begin{eqnarray}
\Theta^j(y_4+1,y_5) &=& e^{\frac{2\pi ij}{n}   } \Theta^{j}(y_4,y_5), 
\nonumber \\ 
\Theta^j(y_4,y_5+1) &=& e^{-\frac{2\pi im}{n} y_4} \Theta^{j+m}(y_4,y_5).  
\end{eqnarray}
Thus, the zero-mode solutions with 
the boundary conditions  (\ref{eq:B-BC-1}) due to non-Abelian Wilson lines 
can be written in terms of $\Theta^j$ of Eq.~(\ref{eq:solution-1}) as 
\begin{eqnarray} \label{eq:zeromode-1}
& & B^j_{pq}(y_4,y_5)=c_{pq}^j \sum_{r=0}^{n-1} e^{2 \pi
  i(m_ap-m_bq)\frac{r}{n}}\Theta^{j+mr},  
\end{eqnarray}
where $j=0,1,...,m-1$.
Here,  $c_{pq}^j$ is a constant normalization, which can be 
determined by the boundary conditions.
Note that the boundary condition  (\ref{eq:B-BC-1}) relates 
$B_{p,q}$ and $B_{p+1,q+1}$.
Thus, there are the series 
\begin{eqnarray}\label{eq:B-series}
B_{p,q} \rightarrow B_{p+1,q+1} \rightarrow \cdots 
\rightarrow B_{p+n,q+n}=B_{p,q}.
\end{eqnarray}
The periodicity of the series is equal to $n$, 
and there are $n$ independent series.
Each of the series has $m$ independent solutions
(\ref{eq:zeromode-1}) and the total number of 
independent zero-modes is equal to $mn$.

We have concentrated on the $B_{+}$ fields.
Similarly, when $mn >0$, 
the $C_-$ fields have the same solutions as $B_+$.
However, the $B_-$ and $C_+$ have no normalizable zero-modes 
for $mn >0$.
On the other hand, when $mn <0$
the $B_-$ and $C_+$ have normalizable zero-modes with the same 
wavefunctions as the above, 
while $B_+$ and $C_-$ have normalizable zero-modes.

We have considered the zero-mode profiles of 
fermionic fields.
If 4D N=1 supersymmetry is preserved, 
the scalar mode has the same zero-mode profiles as its 
fermionic superpartner.

\subsubsection{$n_a \neq n_b$}

Next, we study the model with $n_a \neq n_b$.
In this case, the non-Abelian Wilson lines break
the gauge group $U(N_a)\times U(N_b)$ to 
$U(P_a) \times U(P_b)$, where $P_a=N_a/n_a$ and $P_b=N_b/n_b$.
Similarly to the previous subsection, 
we decompose the fields $B$ as 
\begin{eqnarray}
B= 
\begin{pmatrix}
B_{00} & B_{01} & \cdots & B_{0,n_b-1}\\
B_{10} & B_{11} & \cdots & \\
\cdots \\
B_{n_a-1,0} & B_{n_a-1,1} & \cdots & B_{n_a-1,n_b-1} 
\end{pmatrix}.
\end{eqnarray}
Each of $B_{pq}$ components is  
$(P_a \times P_b)$ matrix-valued fields.
The boundary condition (\ref{eq:B-BC}) due to 
the non-Abelian Wilson lines is written as
\begin{eqnarray}\label{eq:B-BC-1-2}
B_{pq}(y_4+1,y_5) &=& B_{p+1,q+1}(y_4,y_5), \nonumber \\ 
B_{pq}(y_4,y_5+1) &=& e^{-2\pi i (\frac{m_a}{n_a}-\frac{m_b}{n_b}) y_4}
                  e^{-2\pi i(\frac{m_a}{n_a}p-\frac{m_b}{n_a}q)} 
                  B_{p,q}(y_4,y_5),
\end{eqnarray}
where $B_{p+n_a,q}=B_{p,q+n_b}=B_{p,q}$.
This boundary condition relates 
$B_{p,q}$ and $B_{p+1,q+1}$.
Then, similarly to (\ref{eq:B-series}), there are the following series 
\begin{eqnarray}\label{eq:B-series-2}
B_{p,q} \rightarrow B_{p+1,q+1} \rightarrow \cdots 
\rightarrow B_{p+Q_{ab},q+Q_{ab}}=B_{p,q}.
\end{eqnarray}
Here, the periodicity of the series is obtained by 
$Q_{ab} \equiv {\rm l.c.m.}(n_a,n_b)$,\footnote{
Here, l.c.m. denotes the least common multiple.} 
and the number of independent series is equal to 
 $k_{ab} \equiv {\rm g.c.d.}(n_a,n_b)$.
Obviously, there is  the relation, 
$Q_{ab}=\frac{n_a n_b}{k_{ab}}$. 
The above boundary condition (\ref{eq:B-BC-1-2}) 
leads to the boundary condition,
\begin{eqnarray}\label{eq:B-BC-Q}
B_{pq}(y_4+Q_{ab},y_5) &=& B_{pq}(y_4,y_5), \nonumber \\ 
B_{pq}(y_4,y_5+Q_{ab}) &=& e^{-\frac{2\pi i}{k_{ab}} I_{ab} y_4} 
B_{p,q}(y_4,y_5) .
\end{eqnarray}
Here we have defined 'the intersection number' 
$I_{ab} \equiv n_bm_a-n_am_b$ analogous to intersecting brane models.
There are $S_{ab}=\frac{n_an_b}{k^2_{ab}}I_{ab}$ independent zero-mode 
solutions, which satisfy the boundary condition (\ref{eq:B-BC-Q}).
Those functions are obtained as 
\begin{eqnarray}\label{eq:solution-2}
\Theta^{j}(y_4,y_5)
&=&
\sum_n e^{- \pi S_{ab}(n+\frac{j}{S_{ab}})^2 + \frac{2\pi i S_{ab}}{Q_{ab}}
(n+\frac{j}{S_{ab}})y_4-\frac{\pi S_{ab}}{Q_{ab}^2}y_5^2
-2\pi \frac{S_{ab}}{Q_{ab}}(n+\frac{j}{S_{ab}})y_5} \nonumber \\
&=&
e^{-\frac{\pi S_{ab}}{Q_{ab}^2}y_5^2}
\jtheta{\frac{j}{S_{ab}} \\ 0}\left((S_{ab}/Q_{ab})z,S_{ab}\tau\right),
\end{eqnarray}
where $\tau=i$.
These wavefunctions satisfy the following boundary conditions,
\begin{eqnarray}
\Theta^j (y_4+1,y_5) &=& e^{2\pi i \frac{k_{ab}}{n_a n_b}j}
\Theta^j(y_4,y_5) , \nonumber \\
\Theta^j (y_4,y_5+1) &=&
e^{2\pi i(\frac{m_a}{n_a}-\frac{m_b}{n_b})y_4}
\Theta^{j-\frac{I_{ab}}{k_{ab}}}(y_4,y_5) .
\end{eqnarray}
Thus, the zero-mode wavefunctions, which satisfy the 
boundary conditions (\ref{eq:B-BC-1-2}), are obtained as 
\begin{eqnarray}\label{eq:B-wf-2}
B_{pq}^j(y_4,y_5)
= c_{pq}^j 
\sum_{r=0}^{Q_{ab}-1}
e^{2\pi i(\frac{m_a}{n_a} p-\frac{m_b}{n_b}q) r}
\Theta^{j+\frac{I_{ab}}{k_{ab}}r}(y_4,y_5) ,
\end{eqnarray}
where $j =0,1, \cdots, \frac{I_{ab}}{k_{ab}}-1$.
Hence, the number of the independent zero-modes 
is equal to $M_{ab}=\frac{S_{ab}}{Q_{ab}}= \frac{I_{ab}}{k_{ab}}$
in each of the series (\ref{eq:B-series-2}), 
and there are the $k_{ab}$ independent series.
Thus, the total number of zero-modes is equal to 
$M_{ab} k_{ab} = I_{ab}$.

As an illustrating example, we consider the model with 
$n_a=2, n_b=4$ and $m_a=m_b=3$.
Then, we have $k_{ab}={\rm g.c.d.}(n_a,n_b)=2 \ne 1$, 
$Q_{ab}=4$, $S_{ab}=12$ and $I_{ab}=6$.
We decompose the  bi-fundamental fields $B$ with  
the $2\times 4$ matrix entries as
\begin{eqnarray}
B = 
\begin{pmatrix}
B_{00} & B_{01} & B_{02} & B_{03} \\
B_{10} & B_{11} & B_{12} & B_{13}
\end{pmatrix}.
\end{eqnarray}
{}From the wave function formula in Eq.~(\ref{eq:B-wf-2}),
one obtains the three independent solutions labeled by $j=0,1,2$ for each
component of $B_{pq}$ and 
these are represented by linear combinations of $\Theta^i$ 
in Eq.~(\ref{eq:solution-2}).
For example, the $B_{00}$ and $B_{01}$ fields are 
\begin{eqnarray}
B^j_{00} &=& 
\Theta^j +\Theta^{j+3}+\Theta^{j+6} + \Theta^{j+9}, \nonumber \\
B^j_{01} &=& 
\Theta^j + e^{-\frac{3\pi i}{2}} \Theta^{j+3}
+e^{-3\pi i} \Theta^{j+6} + e^{-\frac{9\pi i}{2}} \Theta^{j+9}.
\end{eqnarray}
Obviously, the $y_4$-direction boundary condition can connect some of 
$B_{p,q}$ components follows 
\begin{eqnarray}
B_{00} \to B_{11} \to B_{02} \to B_{13} \to B_{00}, \\
B_{01} \to B_{12} \to B_{03} \to B_{10} \to B_{01}.
\end{eqnarray}
Since there are $k_{ab}=2$ independent series, 
there are $I_{ab}=6$ zero mode solutions in this background.

\subsection{Another representation of solutions}

In the previous section, we have presented 
solutions in terms of the  $\Theta^j$ functions.
However, by using the properties of the theta function, 
one can represent the wave functions (\ref{eq:zeromode-1}) 
and (\ref{eq:B-wf-2}) as a single theta 
function as  
\begin{eqnarray}\label{eq:B-wf-3}
B_{pq}^j(y_4,y_5)
=
C_{p,q}^j
e^{-\pi \tilde{I}_{ab}y_5^2}
\times 
\jtheta{\frac{j}{M_{ab}} \\ 0}
\left(\tilde{I}_{ab} z+ 
\left(\frac{m_a}{n_a}p-\frac{m_b}{n_b}q \right),\tilde{I}_{ab} \tau \right),
\end{eqnarray}
where $\tilde I_{ab} = I_{ab}/n_an_b$.
The constant $C_{p,q}^j$ can be determined by the 
boundary conditions.
The net number of zero-mode multiplicity for each of the series 
is given by $M_{ab}=I_{ab}/k_{ab}$.
Therefore the wave functions $B_{pq}^{j'}(y_4,y_5)$ with $j'=j+M_{ab}$
should be equal to $B_{pq}^j(y_4,y_5)$.
Furthermore we impose the $B_{p+n_a,q}^j=B_{p,q+n_b}^j=B_{p,q}^j$ and 
we have twist boundary condition $B_{pq}^j(y_4+1,y_5)=B_{p+1,q+1}^j(y_4,y_5)$.
Then these conditions imply the following constraint for the
coefficients of $C_{pq}^j$ as
\begin{eqnarray}
e^{2\pi ij\frac{m_a}{M_{ab}} }C_{p+n_a,q}^{j}
=
e^{-2\pi ij\frac{m_b}{M_{ab}} }C_{p,q+n_b}^{j}
=
C_{pq}^j, \\
C_{p+1,q+1}^j=C_{p,q}^j, \quad
C_{p,q}^{j+M_{ab}}=C_{p,q}^j.  
\end{eqnarray}
We start with a certain element, e.g. $C_{0,0}^0$,
Then we fix other elements by using the above relations.
If there are still unrelated elements, 
we start with one of those elements and repeat 
the procedure recursively again.

In general, their solutions for $C^j_{pq}$ can not be determined uniquely.
Only in specific cases, we can write  $C^j_{pq}$ by a simple form.
For example, when $\frac{m_a}{M_{ab}}, \frac{m_b}{M_{ab}}=\rm{integer}$, 
$C_{pq}^j$ is reduced to $C_{pq}^j=\rm{const.}$, 
i.e. independent of $p,q$ and $i$.
Most of the following models correspond to this case.
As another example for a simple form, 
if we can find a certain integer $L$, 
which satisfies 
\begin{eqnarray}
L= \frac{M_{ab}l_a-m_a}{n_a}=-\frac{M_{ab}l_b+m_b}{n_b},
\end{eqnarray}
where $l_a$ and $l_b$ are also integers, 
then we can write $C^j_{pq}$ by the following form
\begin{eqnarray}
C_{pq}^{j}=
e^{ 2\pi i j\frac{L}{M_{ab}}(p-q)}. 
\end{eqnarray}
Then the forms of wave functions would become simple as
\begin{eqnarray}
B_{pq}^j(y_4,y_5)
&=&
e^{ 2\pi i j\frac{L}{M_{ab}}(p-q)} 
e^{-\pi \tilde{I}_{ab} y_5^2} \times
\jtheta{\frac{j}{M_{ab}} \\ 0}
\left( \tilde{I}_{ab} z+ 
\left(\frac{m_a}{n_a}p-\frac{m_b}{n_b}q \right), \tilde{I}_{ab} \tau \right),
\end{eqnarray}
up to a normalization factor.
However this expression is only valid if there exists such an integer $L$ 
satisfying the relations.
At any rate, in generic case we determine $C^j_{pq}$ 
recursively as mentioned above.

So far, we have considered the simple $T^2$, 
where $y_4$ and $ y_5$ are identified as 
  $y_4 \sim y_4+1$ and $ y_5 \sim y_5 +1$.
Similarly, we can study the torus compactfication 
with an arbitrary value of the complex structure modulus $\tau$, 
although we have fixed $\tau =i$ in the above analysis.
Then, we obtain zero-mode wavefunctions 
similar to Eq.~(\ref{eq:B-wf-3})
for an arbitrary value of $\tau$ as Eq.~(\ref{eq:B-wf-3}).
We also replace $z=y_4+iy_5$ in the theta function by $z=y_4+\tau y_5$.

\section{Yukawa couplings}

Here, we study Yukawa couplings.
Let us consider the following form of the magnetic fluxes,
\begin{equation} 
F = 
\begin{pmatrix}
\frac{m_a}{n_a} {\bf 1}_{N_a} & & \\
& \frac{m_b}{n_b} {\bf 1}_{N_b} & \\
& & \frac{m_c}{n_c} {\bf 1}_{N_c} 
\end{pmatrix},
\end{equation}
and non-Abelian Wilson lines similar to (\ref{eq:nA-WL-1}).
Then, there are three types of matter fields, 
$(N_a,\overline N_b)$, $(N_b,\overline N_c)$, 
$(N_c,\overline N_a)$ and their conjugates 
under $U(N_a) \times U(N_b) \times U(N_c)$, 
although they break to $U(P_a) \times U(P_b) \times U(P_c)$
by non-Abelian Wilson lines.
We consider the case with $\frac{m_a}{n_a}-\frac{m_b}{n_b}>0$, 
$\frac{m_b}{n_b}-\frac{m_c}{n_c}>0$
and  $\frac{m_a}{n_a}-\frac{m_c}{n_c}>0$.
Then, the three types of matter fields 
whose wavefunctions are denoted by $\Psi^{j,M_1}$, 
$\Psi^{k,M_2}$ and $(\Psi^{l,M_3})^*$, 
appear in the 
following off-diagonal elements,
\begin{equation} 
\begin{pmatrix}
{\rm{const}} & \Psi^{j,M_1} & \\
& {\rm{const}} &  \Psi^{k,M_2} \\
(\Psi^{l,M_3})^\dagger & & {\rm{const}} 
\end{pmatrix}, 
\end{equation}
where $M_1= M_{ab}$, $M_2=M_{bc}$ and $M_3=M_{ac}$ for simplicity.
We use the same indices for $Q_{ab}$ and others, i.e. 
$Q_1= Q_{ab}$, $Q_2=Q_{bc}$ and $Q_3=Q_{ac}$. 
As already explained, in the background with fractional fluxes 
and non-Abelian Wilson lines, their fields are the matrix valued 
wave functions.
The Yukawa coupling can be calculated by computing the following overlap 
integral of zero-modes in the $(y_4,y_5)$ compact space,
\begin{eqnarray}\label{eq:y-ijk-1}
y^{jkl}_{1,pqr} &=&
 \int_0^1 dy_4\int_0^1 dy_5 
[\Psi^{j,M_1}_{pq}\Psi^{k,M_2}_{qr}(\Psi^{l,M_3}_{rp})^* ] .
\end{eqnarray}
That is, the Yukawa coupling $Y^{ijk}$ in 4D effective theory is obtained as 
their products on $(T^2)^n$, i.e. 
$Y^{ijk} = g_D\prod_{d=1}^{n/2} y^{ijk}_d$, where 
$y_d^{ijk}$ denotes the overlap integral similar to Eq.~(\ref{eq:y-ijk-1}) 
for the $d$-th torus $(T^2)$ and  
$g_D$ is the D-dimensional gauge coupling.
{}From this structure, one can see that the allowed couplings 
are restricted.
In order to see it,  we introduce the following parameters as
$k_1={\rm g.c.d.}(n_a,n_b)$, 
$k_2={\rm g.c.d.}(n_b,n_c)$, 
$k_3={\rm g.c.d.}(n_a,n_c)$
and $K={\rm g.c.d.}(k_1,k_2,k_3)={\rm g.c.d.}(n_a,n_b,n_c)$. 
Then the parameter  $K$ determines the allowed couplings of Yukawa interactions.
If $K=1$, all of possible combinations $(p,q,r)$ appear 
in Eq.~(\ref{eq:y-ijk-1}).
However, if 
$K \ne 1$, only restricted combinations of  $(p,q,r)$ appear 
in Eq.~(\ref{eq:y-ijk-1}), but not all combinations.
That is, the couplings are restricted by the $Z_K$ symmetry.
Indeed, allowed combinations of $(p,q,r)$ are 
controlled by the gauge invariance 
before the gauge symmetry breaking.
This $Z_K$ symmetry is unbroken symmetry in the 
original gauge symmetry.


Now, let us consider the following summation of wavefunction 
products, 
\begin{eqnarray}
I_{pqr}^{jkl} &=&
\Psi^j_{pq} \Psi^k_{qr} (\Psi^l_{rp})^* +\Psi^j_{p+1,q+1}
\Psi^k_{q+1,r+1} (\Psi^l_{r+1,p+1} )^*
\nonumber \\
&& 
+\cdots+ 
\Psi^j_{p+Q-1,q+Q-1} \Psi^k_{q+Q-1,r+Q-1} (\Psi^l_{r+Q-1,p+Q-1})^*,
\end{eqnarray}
where $Q={\rm l.c.m.}(Q_1,Q_2,Q_3)$.
One can represent $Q$ as $Q=Q_1 q_1=Q_2q_2=Q_3 q_3$.
To compute the integral it is useful to represent the wavefunctions
as follows
\begin{eqnarray}
\tilde{\Psi}^{j',M_1'}_{pq}(y_4,y_5)
&=&
C_{pq}^{j'} 
e^{-\pi \frac{M_1'}{Q}y_5^2}
\jtheta{\frac{j'}{M_1'} \\ 0}
\left( \frac{M_1'}{Q }z+ 
\left(\frac{m_a}{n_a}p-\frac{m_b}{n_b}q \right),\frac{M_1'}{Q} \tau \right),
\nonumber \\
\tilde{\Psi}^{k',M_2'}_{qr}(y_4,y_5)
&=&
C_{qr}^{k'} 
e^{-\pi \frac{M_2'}{Q}y_5^2}
\jtheta{\frac{k'}{M_2'} \\ 0}
\left( \frac{M_2'}{Q} z+ 
\left(\frac{m_b}{n_b}q-\frac{m_c}{n_c}r \right),\frac{M_2'}{Q} \tau \right),
\\
\tilde{\Psi}^{l',M_3'}_{pr}(y_4,y_5)
&=&
C_{pr}^{l'} 
e^{-\pi \frac{M_3'}{Q}y_5^2}
\jtheta{\frac{l'}{M_3'} \\ 0}
\left( \frac{M_3'}{Q} z+ 
\left(\frac{m_a}{n_a}p-\frac{m_c}{n_c}r \right),\frac{M_3'}{Q} \tau \right),
\nonumber 
\end{eqnarray}
where $j'=q_1 j$, $k'=q_2 k$, $l'=q_3 l$ and $M_i'=q_i M_i$, ($i=1,2,3$).
Here the relation $M_1'+M_2'=M_3'$ holds. 
By using the product property of the theta function,
the product of $\Psi^{j,M_1}\Psi^{k,M_2}$ is represented by the sum of
the theta functions as
\begin{eqnarray}
\tilde{\Psi}^{j,M_1}_{pq}\tilde{\Psi}^{k,M_2}_{qr}
=
C_{pq}^{j'} C_{qr}^{k'} 
e^{\pi \frac{M_3'}{Q}y_5^2}
\sum_{m\in Z_{M_3'}} 
\jtheta{\frac{j'+k'+M_1'm}{M_3'}}
\left(\frac{M_3'}{Q} z+ 
\left(\frac{m_a}{n_a}p-\frac{m_c}{n_c}r \right),\frac{M_3'}{Q} \tau \right) 
\nonumber \\
\times 
\jtheta{\frac{M_2'j'-M_1'k'+M_1'M_2'm}{M_1'M_2'M_3'}}
\left( \frac{m_a}{n_a}M_2'p-\frac{m_b}{n_b}M_2'q
      -\frac{m_b}{n_b}M_1'q+\frac{m_c}{n_c}M_1'r ,
\frac{M_1'M_2'M_3'}{Q} \tau \right).
\end{eqnarray}

Here one can use the properties of boundary conditions for non-Abelian
Wilson lines. Using the property of
$\Psi_{p,q}(y_4+1,y_5)=\Psi_{p+1,q+1}(y_4,y_5)$ 
we find 
\begin{eqnarray}
\int_0^1 dy_4 \int_0^1 dy_5 I^{ijk}_{pqr} =
\int_0^Q dy_4 \int_0^1 dy_5 \Psi^i_{pq} \Psi^j_{qr} (\Psi^k_{rp})^* .
\end{eqnarray}
Therefore we can obtain the analytic form of Yukawa couplings
and flavor structures similar to the case with Abelian Wilson lines.   
By using the orthogonal condition for the matrix valued wave functions 
(see Appendix A) as
\begin{eqnarray}
\int_0^Q dy_4 \int_0^1 dy_5 \Psi^{j,M_1}_{pq} (\Psi^{k,M_1}_{pq})^\dagger
=\delta_{j,k},
\end{eqnarray}
one can lead to the following form of Yukawa couplings
\begin{eqnarray}
& & \int_0^Q dy_4 \int_0^1 dy_5 \Psi^i_{pq} \Psi^j_{qr} (\Psi^k_{rp})^*
= 
Q\sqrt{\frac{Q}{2M_3'}}
\sum_{m\in Z_{M_3'}}\delta_{j'+k'+M_1'm, ~l' (\rm{mod} {M_3'})}
\nonumber \\
& &~~~~ \times
\jtheta{\frac{M_2'j'-M_1'k'+M_1'M_2'm}{M_1'M_2'(M_3')} \\ 0}
\left(  Q\left(
\frac{m_a}{n_a}\tilde{I}_{bc}p + \frac{m_b}{n_b}\tilde{I}_{ca}q 
+ \frac{m_c}{n_c}\tilde{I}_{ab}r  \right)  ,\frac{M_1'M_2'M_3'}{Q} \tau \right).
\end{eqnarray}
up to the factor $N_{M_1}N_{M_2}N_{M_3}^* C_{pq}^j C_{qr}^k (C_{pr}^l)^*$.
Here, the Kronecker delta $\delta_{j'+k'+M_1'm, ~l' (\rm{mod} {M_3'})}$
leads to the  coupling selection rule 
\begin{eqnarray}
j'+k'+M_1'm=l'\mod{M_3'},
\end{eqnarray}
where $m=0,1,...,M_3'-1$.
When $g={\rm g.c.d.}(M_1',M_2',M_3')={\rm g.c.d.}(M_1,M_2,M_3)$, 
the coupling selection rule is given by 
\begin{eqnarray}\label{eq:Zg}
j'+k'=l' \mod{g}.
\end{eqnarray}
That means that we can assign $Z_g$ charges to 
all of zero-modes.\footnote{See
  Refs.~\cite{Cremades:2003qj,Higaki:2005ie} 
for a similar selection rule in intersecting D-brane models.}

Here we study again the $Z_K$ symmetry, which we showed.
The total number of multiplicity of $\Psi_{ab}$ is nothing but $|I_{ab}|$, 
and it is represented by two parameters of $k_{ab}$ and $M_{ab}$ as 
 $I_{ab}=k_{ab}M_{ab}$.
If $K={\rm g.c.d.}(k_{ab},k_{bc},k_{ca}) \ne 1$, 
they are divided to $K$ types of zero-modes and 
distinguished by labeling the component of each matrix.
We introduce such a kind of flavor indices as $\tilde{j}, \tilde{k}$
and $\tilde{l}$ for $ab$-, $bc$-, $ca$-sectors, respectively.
We define the relation between the flavor labeled by $\tilde{j}$ 
and the component of matrix $p,q$ as $\tilde{j}=p-q \mod{k_1}$.
Similarly the other sectors are also defined as
$\tilde{k}=q-r \mod{k_2}$  and $\tilde{l}=p-r \mod{k_3}$. 
Since the allowed couplings must be gauge invariant, 
there is the coupling selection rule for this kind of flavor indices, 
which is given by 
\begin{eqnarray}\label{eq:ZK}
\tilde{j}+\tilde{k}=\tilde{l} \mod{K}.
\end{eqnarray}
This is because the Yukawa couplings are 
restricted in the trace of the matrix.
Therefore we find two types of coupling selection rules, 
i.e. the $Z_g$ and $Z_K$ symmetries.

We can extend the computation of 3-point couplings to 
higher order couplings.
For example, in appendix B, we show the computation of 4-point couplings.

\section{Non-Abelian discrete flavor symmetry}

Here, we study the non-Abelian flavor symmetries, 
which can appear in our models.

\subsection{The case with $M_i \ne 1$ and $k_i=1$ }

First, we consider the models with $k_1=k_2=k_3=1$.
Then, the number of zero-modes are given by $|I_{ab}|=M_1$,
$|I_{bc}|=M_2$ and $|I_{ca}|=M_3$.
We consider the models with $g={\rm g.c.d.}(M_1,M_2,M_3)\ne 1$.
The Yukawa couplings do not depend on the matrix components  
$(p, q,r)$, and are reduced to the following form 
\begin{eqnarray}
\int_0^Q dy_4 \int_0^1 dy_5 \Psi^i_{pq} \Psi^j_{qr} \Psi^k_{rp}
= 
N_{M_1}N_{M_2}N_{M_3}^* 
Q\sqrt{\frac{Q}{2M_3'}}
\sum_{m\in Z_{M_3'}}\delta_{j'+k'+M_1'm, l' (\rm{mod} {M_3'})}
\nonumber \\
\times
\jtheta{\frac{M_2'j'-M_1'k'+M_1'M_2'm}{M_1'M_2'(M_3')} \\ 0}
\left(  0 ,M_1'M_2'M_3'/Q \tau \right),
\end{eqnarray}
where we have taken simply $p=q=r=0$ and 
the phase factor like $C_{pq}^j$ disappears.
This form is nothing but the case with integer fluxes and 
without non-Abelian Wilson lines.
In this types of Yukawa couplings,  
4D effective theory has another flavor symmetry called 
by the shift symmetry, which 
corresponds to the transformations of flavor indices as
\begin{eqnarray}\label{eq:permutation-g}
& & j' \to j'+M'_1/g, \nonumber \\ 
& & k' \to k'+M'_2/g, \\
& & l' \to l'+M'_3/g, \nonumber
\end{eqnarray}
simultaneously. 
Under this transformation, Yukawa couplings are invariant. 
There is also coupling selection rule as shown in the previous section
given by the $Z_g$ symmetry (\ref{eq:Zg}).
Then, they form the non-Abelian discrete flavor symmetries 
as the same as the 
case without non-Abelian Wilson lines.

For simplicity, suppose that $M'_1=g$.
Then, there are $g$ zero-modes of  $\Psi^{j',M'_1}$.
The selection rule (\ref{eq:Zg}) means that 4D effective theory is 
symmetric under the  $Z_g$ transformation, which 
acts on $\Psi^{j',g}$ as 
$Z \Psi^{j',g}$, where
\begin{eqnarray}
Z = \left(
\begin{array}{ccccc}
1 & & & & \\
  & \rho & & & \\
  & & \rho^2 & & \\
  & &   & \ddots & \\
  &  &  &    & \rho^{g-1} 
\end{array}
\right),
\end{eqnarray}
and $\rho = e^{2\pi i /g}$.
Furthermore, 
the effective theory has another symmetry (\ref{eq:permutation-g}).
That  can be written as cyclic permutations on  $\Psi^{j',g}$, 
\begin{equation}\label{eq:permutation}
\Psi^{j',g} \rightarrow \Psi^{j'+1,g} .
\end{equation}
That is nothing but a change of ordering and also 
has a geometrical meaning as a discrete shift 
of the origin, $z=0 \rightarrow z=-\frac{1}{g}$.
This symmetry also generates another $Z_g$ symmetry, 
which we denote by $Z_g^{(C)}$ and its generator is 
represented as 
\begin{eqnarray}\label{eq:C}
C = \left(
\begin{array}{cccccc}
0 & 1& 0 & 0 & \cdots & 0 \\
0  & 0 &1 & 0 & \cdots & 0\\
  &    &  & &\ddots & \\
1  &  0  & 0 &  & \cdots   & 0 
\end{array}
\right),
\end{eqnarray}
on $\Psi^{j',g}$.
These generators, $Z$ and $C$, do not commute each other,
i.e., 
\begin{equation}
CZ = \rho ZC .
\end{equation}
Then, the flavor symmetry corresponds to the closed algebra 
including $Z$ and $C$.
Diagonal matrices in this closed algebra are written as 
$Z^n(Z')^m$, 
where $Z'$ is the generator of another $Z'_g$  
written as 
\begin{eqnarray}
Z' = \left(
\begin{array}{ccc}
\rho & &  \\
    & \ddots & \\
    &    & \rho 
\end{array}
\right),
\end{eqnarray}
on $\Psi^{j',g}$.
Hence, these would generate the non-Abelian flavor symmetry 
$(Z_g \times Z'_g)\rtimes Z_g^{(C)}$, since 
$Z_g \times Z'_g$ is a normal subgroup.
These discrete flavor groups would include $g^3$ elements totally.

For example, for $g=2$ and 3 these flavor symmetries are given as 
$Z_2 \rtimes Z_2 =D_4$ 
and $(Z_3 \times Z_3) \rtimes Z_3 = \Delta(27)$, respectively.
Then, the fields $\Psi^{j',g}$ correspond to 
${\bf 2}$ of $D_4$ and ${\bf 3}$ of $\Delta (27)$, as shown in 
Tables \ref{tab:g2} and \ref{tab:g3}, respectively.
When $M'/g$ is an integer larger than $1$, 
the $\Psi^{j',M'}$ fields correspond to other representations.
For smaller values of $M'/g$, the corresponding representations 
are shown in Tables \ref{tab:g2} and \ref{tab:g3}.

However we note that
their multiplets have several types of representation under 
this symmetry.
Because a $Z_g$ charge of fields labeled by $j$ is not 
$j$ but $j'=qj$.
Therefore even if they have same multiplicities ($M_1=M_2$),
their representations may be different from each other.  

\begin{table}[t]
\begin{center}
\begin{tabular}{|c|c|} \hline
$M'$ & Representation of $D_4$ \\ \hline \hline
2 & ${\bf 2}$ \\ 
4 & ${\bf 1}_{++}, \ {\bf 1}_{+-}, \ {\bf 1}_{-+}, \ {\bf 1}_{--}$ \\ 
6 & $3 \times {\bf 2}$\\  \hline
\end{tabular}
\end{center}
\caption{$D_4$ representations of zero-modes in the model with $g=2$.}
\label{tab:g2}
\end{table}

\begin{table}[t]
\begin{center}
\begin{tabular}{|c|c|} \hline
$M'$ & Representation of $\Delta(27)$ \\ \hline \hline
3 & ${\bf 3}$ \\ 
6 & $2 \times {\bar {\bf 3}}$  \\ 
9 & ${\bf 1}_{1}, \ {\bf 1}_{2}, \ {\bf 1}_{3}, \ {\bf 1}_{4}, 
\ {\bf 1}_{5}, \ {\bf 1}_{6}, \ {\bf 1}_{7}, \ {\bf 1}_{8}, 
\ {\bf 1}_{9}$ \\ 
12 & $4 \times {\bf 3}$\\  
15 & $5 \times {\bar {\bf 3}}$  \\ 
18 & $2 \times \{ {\bf 1}_{1}, \ {\bf 1}_{2}, \ {\bf 1}_{3}, \ {\bf 1}_{4}, 
\ {\bf 1}_{5}, \ {\bf 1}_{6}, \ {\bf 1}_{7}, \ {\bf 1}_{8}, 
\ {\bf 1}_{9} \}$ \\ \hline
\end{tabular}
\end{center}
\caption{$\Delta(27)$ representations of zero-modes in the model with $g=3$.}
\label{tab:g3}
\end{table}

\subsection{The case with $M_i = 1$ and $k \ne 1$ }

Next, we consider the models with $M_i = 1$ and $k \ne 1$.
In this case, we also find flavor structures similar to  
the case without non-Abelian Wilson lines.  
Suppose all the components of zero modes are given by 
$|I_{ab}|=k_1$, $|I_{bc}|=k_2$ and $|I_{ca}|=k_3$.
Then it is possible to take phase factors for each of wave functions 
$C^j_{pq}=1$.
We commonly use $K={\rm g.c.d.}(k_1,k_2,k_3)$.
The Yukawa couplings depend on the indices $p,q$ and $r$
only through a combination $\theta_{pqr}$ given by 
\begin{eqnarray}
\theta_{pqr}&=&
Q\left( \frac{m_a}{n_a}\tilde{I}_{bc}p
+ \frac{m_b}{n_b}\tilde{I}_{ca}q
+ \frac{m_c}{n_c}\tilde{I}_{ab}r
\right) \nonumber \\
&=&
Q\left( \frac{m_a}{n_a}\tilde{I}_{bc}(\tilde{j}+n_1 k_1)
- \frac{m_c}{n_c}\tilde{I}_{ab}(\tilde{k}+n_2 k_2)
\right) ,
\end{eqnarray}
where we have used the relations $p-q=n_1 k_1+\tilde{j}$ and 
$l-r=n_2 k_2+\tilde{k}$ with $n_1, n_2 \in Z$.
We find that the Yukawa couplings are invariant under the following
transformation as
\begin{eqnarray}\label{eq:shift-2}
\tilde{j} &\to& \tilde{j} + \frac{m_c I_{ab}}{K}, \nonumber \\
\tilde{k} &\to& \tilde{k} + \frac{m_a I_{bc}}{K}, \\
\tilde{l} &\to& \tilde{l} + \frac{m_b I_{ac}}{K}. \nonumber 
\end{eqnarray}
It is obvious that this transformation is the permutation of flavor
index with order $K$.
Therefore we have two symmetries: one is 
the discrete $Z_K$ symmetry comes from the coupling selection rule and 
another is this shift symmetry.
By combining these two symmetries, it become the same 
non-Abelian discrete flavor symmetry 
as the case without Non-Abelian Wilson-lines.
That is, these flavor symmetries are given as 
$Z_2 \rtimes Z_2 =D_4$ for $K=2$, 
$(Z_3 \times Z_3) \rtimes Z_3 = \Delta(27)$ for $K=3$ and 
$(Z_K \times Z_K) \rtimes Z_K $ for generic $K$.

We have two aspects of flavor structures which are characterized
by the parameters $M,K$.
In the latter case, the origin of flavor symmetry is  the gauge symmetry.
The background breaks the continuous gauge symmetry, 
but discrete symmetry remains as the flavor symmetry.
In the former case, the flavor would not directly originated from 
the gauge symmetry.
However, T-duals of both cases would correspond to 
similar intersecting $D$-brane models, where 
$n_a$ and $m_a$ have almost the same meaning, that is, 
winding numbers of $D$-branes for different directions.
Thus, these two pictures of flavor symmetries are 
related with each other by T-duality through the intersecting 
$D$-brane picture.

So far, we have considered the models with $M_i =1$ and $K \ne 1$ 
and found the flavor symmetry $(Z_K \times Z_K) \rtimes Z_K$.
Here we comment on generic case with  $M \ne 1$ and $K \ne 1$.
Even in such a case, the selection rules due to $Z_g$ and $Z_K$ 
symmetries hold exact.
However, the general formula of Yukawa couplings depend on both the indices
$j$ and  $\tilde{j}$. 
Then, 4D effective Lagrangian is not always invariant under 
the above (independent) shift transformations 
(\ref{eq:permutation-g}) and (\ref{eq:shift-2}).

\subsection{Illustrating examples}

We show two illustrating examples.
We concentrate on  only the $T^2$ torus.
The first example is the model with $(I_1,I_2,I_3)=(2,4,2)$.
The background magnetic flux is taken as 
\begin{eqnarray}
F=
2\pi 
\begin{pmatrix}
\frac{1}{2} {\bf 1}_{N_a} & & \\ 
& \frac{3}{8} {\bf 1}_{N_b} & \\ 
& & \frac{1}{4} {\bf 1}_{N_c} \\ 
\end{pmatrix}.
\end{eqnarray}
Then the appearing chiral matters are denoted by 
\begin{eqnarray} 
\lambda=
\begin{pmatrix}
\rm{const} & L_{pq}^{j,M_1=1} & \\
& \rm{const} &  R_{qr}^{k,M_2=1} \\
H_{rp}^{l,M_3=1} & & \rm{const} 
\end{pmatrix}, 
\end{eqnarray}
where $p=$0, 1, $q=$0, 1, ..., 7 and $r=$0, 1, 2, 3.
The wave functions are represented by following theta functions as
\begin{eqnarray} 
L_{pq}^j(y)
&=&
N_{M_1}e^{-\pi/8 y^2}
\jtheta{0 \\ 0}(z/8+(1/2p-3/8q),\tau/8),  \nonumber \\
R_{qr}^k(y)
&=&
N_{M_2}e^{-\pi/8 y^2}
\jtheta{0 \\ 0}(z/8+(3/8q-1/4r),\tau/8),  \nonumber \\
H_{pr}^l(y)
&=&
N_{M_3}e^{-\pi/4 y^2}
\jtheta{0 \\ 0}(z/4+(1/2p-1/4r),2\tau/8)  ,
\end{eqnarray}
where we take $j=k=l=0$. 
The several parameters are also given by these fluxes.
We have $k_1=2$, $k_2=4$, $k_3=2$ and  
$K={\rm g.c.d.}(k_1,k_2,k_3)=2$. 
The gauge invariant 3-point couplings are divided to four types
of Yukawa couplings shown below 
\begin{eqnarray} 
\mathcal{L}
&=&
\mathcal{L}_{000}+\mathcal{L}_{010}+\mathcal{L}_{001}+\mathcal{L}_{100} , \nonumber \\
\mathcal{L}_{000} &=&
 L_{00}R_{00}H_{00}^\dagger+L_{11}R_{11}H_{11}^\dagger
+L_{02}R_{22}H_{02}^\dagger+L_{13}R_{33}H_{13}^\dagger 
\nonumber \\
&&
+L_{04}R_{40}H_{00}^\dagger+L_{15}R_{51}H_{11}^\dagger
+L_{06}R_{62}H_{02}^\dagger+L_{17}R_{73}H_{13}^\dagger ,
\nonumber \\
\mathcal{L}_{011} &=&
 L_{00}R_{01}H_{01}^\dagger+L_{11}R_{12}H_{12}^\dagger
+L_{02}R_{23}H_{03}^\dagger+L_{13}R_{30}H_{10}^\dagger
\nonumber \\
&&
+L_{04}R_{41}H_{01}^\dagger+L_{15}R_{52}H_{12}^\dagger
+L_{06}R_{63}H_{03}^\dagger+L_{17}R_{70}H_{10}^\dagger ,
\nonumber \\
\mathcal{L}_{101} &=&
 L_{10}R_{00}H_{10}^\dagger+L_{01}R_{11}H_{01}^\dagger
+L_{12}R_{22}H_{12}^\dagger+L_{03}R_{33}H_{03}^\dagger
\nonumber \\
&&
+L_{14}R_{40}H_{10}^\dagger+L_{05}R_{51}H_{01}^\dagger
+L_{16}R_{62}H_{12}^\dagger+L_{07}R_{73}H_{03}^\dagger ,
\nonumber \\
\mathcal{L}_{110} &=&
 L_{10}R_{01}H_{11}^\dagger+L_{01}R_{12}H_{01}^\dagger
+L_{12}R_{23}H_{13}^\dagger+L_{03}R_{30}H_{00}^\dagger
\nonumber \\
&&
+L_{14}R_{41}H_{11}^\dagger+L_{05}R_{52}H_{02}^\dagger
+L_{16}R_{63}H_{13}^\dagger+L_{07}R_{70}H_{00}^\dagger .
\nonumber 
\end{eqnarray}
As seen in these interaction terms, 
one finds that all the combinations $(p,q,r)$ are not allowed.
This is because it has $K={\rm g.c.d.}(2,4,2)=2$. 
These fields $L,R,H$ are divided to two classes under the discrete
$Z_2$ charge.
For instance, for $R$ fields, 
the flavor index is defined by $\tilde{k}=q-r\mod{4}$.
We assign the $Z_2$ charges as  
\begin{eqnarray} 
Z_2 \ + &:& R^{\tilde{k}=0}, R^{\tilde{k}=2}, \nonumber \\
Z_2 \ - &:& R^{\tilde{k}=1}, R^{\tilde{k}=3},
\end{eqnarray}
and for other fields we also assign the $Z_2$ charges as
\begin{eqnarray} 
Z_2 \ + &:& L^{\tilde{k}=0}, H^{\tilde{k}=0}, \nonumber \\
Z_2 \ - &:& L^{\tilde{k}=1}, H^{\tilde{k}=1}.
\end{eqnarray}
That corresponds to the coupling selection rule as 
$\tilde{j}+\tilde{k}=\tilde{l} \mod{2}$.
The Yukawa couplings $Y^{jkl}_{pqr}$ are obtained after the 
overlap integrals as 
\begin{eqnarray} 
Y_{pqr}^{jkl}
\propto 
\jtheta{0 \\ 0}(1/2p-3/4q+1/4r,\tau/4).
\end{eqnarray}
We also consider about the shift symmetry for this model, i.e.
\begin{eqnarray} 
\tilde{j} &\to& \tilde{j} + \frac{m_c I_{ab}}{K}=\tilde{j}+1 \mod{2}, \nonumber \\
\tilde{k} &\to& \tilde{k} + \frac{m_a I_{ab}}{K}=\tilde{k}+2 \mod{4}, \\
\tilde{l} &\to& \tilde{l} + \frac{m_b I_{ab}}{K}=\tilde{l}+1 \mod{2}. \nonumber
\end{eqnarray}
As shown in the previous section, the Yukawa couplings are 
invariant under this transformation.
These two operators make the $D_4=Z_2 \rtimes Z_2$ discrete flavor symmetry. 
One can understand the representation for each field under $D_4$ symmetry.
As an analysis similar to the previous section, 
one can find that $L$ and $R$ correspond to doublets and $H$ fields become 
four non-trivial singlets under $D_4$ symmetry.

As another example, we consider the model with 
$(I_1,I_2,I_3)=(3,3,3)$, which is not realized by only integer fluxes. 
We choose fluxes as 
\begin{eqnarray}
F=
2\pi 
\begin{pmatrix}
3 {\bf 1}_{N_a} & & \\ 
& \frac{3}{2} {\bf 1}_{N_b} & \\ 
& & 0 {\bf 1}_{N_c} \\ 
\end{pmatrix}.
\end{eqnarray}
Then the appearing chiral matter fields are denoted as follows,
\begin{eqnarray} 
\lambda=
\begin{pmatrix}
\rm{const} & L_{0p}^{j,M_1=3} & \\
& \rm{const} &  R_{q0}^{k,M_2=3} \\
H_{00}^{l,M_3=3} & & \rm{const} 
\end{pmatrix}, 
\end{eqnarray}
where $p,q=0,1$.
This model has $k_i=1$ and $Q_1=2, Q_2=2, Q_3=1, Q=2$
$(j'=j,k'=k,l'=2l)$. 
The gauge invariant 3-point couplings are given as 
\begin{eqnarray} 
\mathcal{L}
&=&
{\rm{tr}}L_{pq}R_{qr}H_{pr}^\dagger \nonumber \\
&=&
L_{00}R_{00}H_{00}^\dagger+L_{01}R_{10}H_{00}^\dagger . 
\end{eqnarray}
The Yukawa couplings $Y^{jkl}_{pqr}$ are calculated by overlap integrals as follows
\begin{eqnarray} 
Y_{pqr}^{jkl}
&=&
\int_0^1 dy_4 \int_0^2 dy_5 L^j_{pq}(y)R^k_{qr}(y)H^l_{rp}(y)^*
\nonumber \\
&\propto& 
\sum_{m \in Z_6} 
\delta_{j'+k'+3m,l'}
\jtheta{\frac{3j'-3k'+9m}{54} \\ 0}(0,27\tau), 
\end{eqnarray}
where we take $p=q=r=0$.
{}From the structure of Kronecker delta,
one can read the selection rule as
\begin{eqnarray} 
&j'+k'+3m=l' \mod{6}& \nonumber \\ 
\to &j+k-2l=0 \mod{3}&.
\end{eqnarray}
Since $g$ is defined by $g={\rm g.c.d.}(M_1',M_2',M_3')=3$,
this model has $\Delta(27) = (Z_3 \times Z_3) \rtimes Z_3$ flavor symmetry.
Here we mention that the charge assignment is different from 
the case with Abelian Wilson line.
For $H$ fields, their $Z_3$ charges are obtained as $l'=2l$, 
and they correspond to the multiplet of $\bar{\bf{3}}$ representations.
Other fields, $L$ and $R$, correspond to $\bf{3}$
representations, and they can couple in the language of 
flavor symmetry. 
Therefore the extension to the non-Abelian Wilson line case causes to 
have more various types of representations and flavor structures.

It is possible to introduce the constant gauge 
potential called by the Abelian Wilson line.
We take the previous model with $(I_1,I_2,I_3)=(3,3,3)$.
We assume $N_a=4$, $N_b=4$, $N_c=2$.
Then the fractional fluxes with non-Abelian Wilson lines 
can reduce the rank of gauge symmetry,
that is, the $U_b(4)$ gauge group breaks to $U_b(2)$ and 
the total gauge symmetry is $U(4)_a\times U(2)_b\times U(2)_c$.
To break the gauge symmetry $U(4)_a\times U(2)_b\times U(2)_c$ 
to the standard-model gauge group, 
Abelian Wilson lines can be introduced.
For example, we can introduce the Abelian Wilson lines in 
$U(4)_a$ along the following direction,
\begin{eqnarray}
\begin{pmatrix}
a_1 {\bf 1}_{3} & \\ 
& a_2 {\bf 1}_{1} \\ 
\end{pmatrix}.
\end{eqnarray}
Then, the gauge group $U(4)_a$ is broken to $U(3)\times U(1)$.
Similarly, we introduce the Abelian Wilson lines in 
$U(2)_c$ along the following direction,
\begin{eqnarray}
\begin{pmatrix}
c_1 {\bf 1}_{1} & \\ 
& c_2 {\bf 1}_{1} \\ 
\end{pmatrix}.
\end{eqnarray}
Then, the gauge group $U(2)_c$ is broken to $U(1)\times U(1)$.
Then the (supersymmetric) standard model with three generations
is realized up to $U(1)$ factors.
We can also introduce the Abelian Wilson line along 
the $U(1)$ direction of $U(2)_b$.
Since the different Wilson line leads to different Yukawa couplings,
that would lead to various flavor structures.
For example, the above model leads to the $\Delta(27)$ flavor 
symmetry in generic values of Wilson lines as studied in 
the previous section.
However, the flavor symmetry is enhanced to 
the $\Delta(54)$ symmetry when Wilson lines vanish.
Thus by choosing the particular choice of Abelian Wilson lines,
we could realize that the flavor symmetry is large like $\Delta(54)$ 
in a subsector, 
e.g. in the lepton sector, but the other sector, e.g. the quark 
sector,  has the smaller flavor symmetry like $\Delta(27)$.\footnote{
Indeed, non-Abelian discrete flavor symmetries such as 
$D_4$, $\Delta(27)$ and $\Delta(54)$ would lead to 
phenomenologically interesting models  
\cite{Grimus,Branco:1983tn,Ishimori:2008uc}.}

\section{Magnetized orbifold background}

We have obtained the explicit wavefunctions on the torus background.
Here, we study about the models on the orbifold background. 
Following Ref.~\cite{Abe:2008fi}, we study the $T^2/Z_2$ orbifold,
which is constructed by dividing $T^2$ by the $Z_2$ projection $z \to -z$.
Furthermore, we require the field projection of periodic or
anti-periodic boundary conditions consistent with the $Z_2$ orbifold,
\begin{eqnarray}
 \Psi(-y_4,-y_5)=P \Psi(y_4,y_5), 
\end{eqnarray}
where $P$ is $+1$ or $-1$.
One can show that the matter wave functions satisfy the following
property
\begin{eqnarray}\label{eq:z2-property} 
\Psi_{pq}^j(-y_4,-y_5)=\Psi_{-p,-q}^{-j}(y_4,y_5).
\end{eqnarray}
For the case with $k=1$, this relation holds, 
because every sector of $(p,q)$ are related by the boundary
conditions, so the labels $(p,q)$ have no meaning.
However, in the $k \ne 1$ case, they have $k \times M$ independent zero-modes 
and we symbolically denote them by $\Psi^{j,\tilde j}$ 
($j=0,1,...,M-1$ and $\tilde j=0,1,...,k-1$).
For example, in the case with $n_a=n_b=3$, we may 
use the following notations
\begin{eqnarray} 
\Psi_{00}^j,\  \Psi_{11}^j,\  \Psi_{22}^j \to  \Psi^{j,\tilde j=0}, \nonumber \\
\Psi_{01}^j,\  \Psi_{12}^j,\  \Psi_{20}^j \to  \Psi^{j,\tilde j=1},  \\
\Psi_{02}^j,\  \Psi_{10}^j,\  \Psi_{21}^j \to  \Psi^{j,\tilde j=2}, \nonumber 
\end{eqnarray}
where $\tilde j= p-q$ mod $K$.
Then, the above property (\ref{eq:z2-property}) can be written as
\begin{eqnarray} 
\Psi^{j,\tilde j}(-y_4,-y_5)=\Psi^{-j,-\tilde j}(y_4,y_5).
\end{eqnarray}
Then the even and odd wave-functions are easily obtained.
For the case with $M=3$, there are $3\times 3$ independent fields 
and they are divided into the following even and odd wave functions
\begin{eqnarray} 
{\rm{even}} &:& \Psi^{0,0},\  
\Psi^{1,0} +\Psi^{2, 0},\  
\Psi^{0, 1}+\Psi^{0, 2},\ 
\Psi^{1,1}+\Psi^{2,2},
\Psi^{2,1}+\Psi^{1,2}, \nonumber \\ 
{\rm{odd}}  &:& \Psi^{1,0}-\Psi^{2,0},\
\Psi^{1,1}-\Psi^{2,2},\  
\Psi^{2,1}-\Psi^{1,2}.
\end{eqnarray}
Note that 
these represent the wave functions e.g. $\Psi_{12}^1+\Psi_{21}^2$ 
by $\Psi^{1,1}+\Psi^{2,2}$. 
As examples, the zero-mode numbers of even and odd wave functions 
for smaller values of $k$ and $M$ 
are shown in Table \ref{tab:orbifold}.

\begin{table}[httbp]
\begin{center}
$k=1$
\begin{tabular}{c|cccccc} 
   $M$ & 1 & 2 & 3 & 4 & 5 & 6 \\ \hline 
even & 1 & 2 & 2 & 3 & 3 & 4 \\
odd  & 0 & 0 & 1 & 1 & 2 & 2 
\end{tabular} \ \ \ 
$k=2$
\begin{tabular}{c|cccccc} 
   $M$ & 1 & 2 & 3 & 4 & 5 & 6 \\ \hline 
even & 2 & 4 & 4 & 6 & 6 & 8 \\
odd  & 0 & 0 & 2 & 2 & 4 & 4 
\end{tabular}
\\ 
$k=3$
\begin{tabular}{c|cccccc} 
   $M$ & 1 & 2 & 3 & 4 & 5 & 6  \\ \hline 
even & 2 & 4 & 5 & 7 & 8 & 10 \\
odd  & 1 & 2 & 4 & 5 & 7 & 8 
\end{tabular} \ \ \ 
$k=4$
\begin{tabular}{c|cccccc} 
   $M$ & 1 & 2 & 3 &  4 &  5 &  6 \\ \hline 
even & 3 & 6 & 7 & 10 & 11 & 14 \\
odd  & 1 & 2 & 5 &  6 &  9 & 10 
\end{tabular}
\end{center}
\caption{The numbers of even and odd zero-modes}
\label{tab:orbifold}
\end{table}

Yukawa couplings as well as higher order couplings 
can be computed on the orbifold background by 
overlap integrals of wavefunctions in a way 
similar to the torus models.

\section{Conclusion}

We have studied the flavor structure of 
4D effective theories, which are derived from 
extra dimensional theories with 
magnetic fluxes and non-Abelian Wilson lines.
We have obtained zero-mode wavefunctions for generic case.
Their Yukawa couplings as well as four-point couplings have been 
computed.
Furthermore, we have also studied non-Abelian flavor symmetries 
and some parts of them are originated from gauge symmetries.
In addition, the orbifold compactification has been discussed.

We have obtained quite rich flavor structure 
compared with the magnetized torus/orbifold models with 
Abelian flavor structures.
For example, there are various ways of model building leading 
to three generation models.
Thus, it would be interesting to apply our analysis 
to phenomenological model building.

\subsection*{Acknowledgement}

H.~A. is supported in part by the Waseda University Grant for 
Special Research Projects No.~2009A-854.
K.-S.~C., T.~K. and H.~O. are supported in part by the Grant-in-Aid for 
Scientific Research No.~20$\cdot$08326, No.~20540266 and
No.~21$\cdot$897 from the 
Ministry of Education, Culture, Sports, Science and Technology of Japan.
T.~K. is also supported in part by the Grant-in-Aid for the Global COE 
Program "The Next Generation of Physics, Spun from Universality and 
Emergence" from the Ministry of Education, Culture,Sports, Science and 
Technology of Japan.

\appendix

\section{Integral of wave functions}

In this appendix we show some calculations on 
integral of wave functions, which are used in section 4. 
The general wave functions are expressed as

\begin{eqnarray}
\Psi^{j,M}_{pq}(y_4,y_5)
&=&
C_{pq}^{j} N_M
e^{-\pi \frac{M'}{Q}y^2_5}
\jtheta{\frac{j'}{M'} \\ 0}
\left( \frac{M'}{Q} z+ 
\left(\frac{m_a}{n_a}p-\frac{m_b}{n_b}q \right),\frac{M'}{Q} \tau \right)
\nonumber \\
\end{eqnarray}
where $Q$ is given by integer satisfying the relation
$Q=k \times {\rm l.c.m.}(n_a,n_b)$, $k \in Z$.
The orthogonal condition implies  
\begin{align}
\int_0^Q dy_4  \int_0^1 dy_5 &  \Psi^{j,M}_{pq} (\Psi^{k,M}_{pq})^\dagger
=
|N_M|^2 C_{pq}^{j} (C_{pq}^{k})^*
\int_0^Q dy_4 
\int_0^1 dy_5 
e^{2\pi M/Q y^2_5}
\nonumber \\
&\times 
\sum_l 
e^{-\pi M'/Q (l+j'/M')^2 }e^{2\pi i\frac{M'}{Q}(l+j'/M') 
\left( y_4+iy_5+ \left( \frac{m_a}{n_a}p-\frac{m_b}{n_b}q \right) \right) }
\nonumber \\
&\times 
\sum_m
e^{-\pi M'/Q (m+j'/M')^2 }e^{-2\pi i\frac{M'}{Q}(m+j'/M') 
\left(y_4-iy_5+ \left(\frac{m_a}{n_a}p-\frac{m_b}{n_b}q \right) \right) }.
\end{align}
Here the integral over $y_4$ is obtained as 
\begin{eqnarray}
\int_0^Q dy_4  
e^{2\pi iy_4 \frac{M'}{Q} \left\{(l+j'/M')- (m+j'/M') \right\} }
&=&
Q \delta_{l, m ({\rm{mod}}\ M)}  \delta_{j, k({\rm{mod}}\ M)}.
\end{eqnarray}
Then, we obtain
\begin{eqnarray}
\int_0^Q dy_4  \int_0^1 dy_5   \Psi^{j,M}_{pq} (\Psi^{k,M}_{pq})^\dagger
&=& Q|N_M|^2
\int_{-\infty}^{\infty} dy_5 e^{-\frac{2\pi M'}{Q}y^2_5} 
\delta_{j, k ({\rm{mod}}\ M)}
\nonumber \\
&=& 
Q \sqrt{\frac{Q}{2M'}} |N_M|^2
\delta_{j, k ({\rm{mod}}\ M)}.
\end{eqnarray}
By this we can fix the normalization.

\section{N-point coupling}

One can calculate the N-point couplings 
by using the addition formula of the theta functions
as in Ref.~\cite{Abe:2009dr}.
For example, we show here the explicit calculation for 
general four point couplings.
We assume that $\tilde{I}_{ab}$, $\tilde{I}_{bc}$, $\tilde{I}_{cd}>0$
and $\tilde{I}_{da}<0$.
Four zero-mode wavefunctions are written as 
\begin{eqnarray}
\psi^{j,M_1}_{pq} &=&
C_{pq}e^{-\pi \frac{M_1'}{Q}y^2_5}
\jtheta{ j'/M_1' \\ 0 }\left( \frac{M_1'}{Q} z +(\frac{m_a}{n_a}p-
\frac{m_b}{n_b}q ), \frac{M_1'}{Q} \tau\right),
\nonumber \\
\psi^{k,M_2}_{qr} &=&
C_{qr}e^{-\pi \frac{M_2'}{Q}y^2_5}
\jtheta{ k'/M_2' \\ 0 }\left(\frac{ M_2'}{Q} z +(\frac{m_b}{n_b}q-
\frac{m_c}{n_c}r ), \frac{M_2'}{Q} \tau\right),
\nonumber \\
\psi^{l,M_3}_{rs} &=&
C_{rs}e^{-\pi \frac{M_3'}{Q}y^2_5}
\jtheta{ l'/M_3' \\ 0 }\left( \frac{M_3'}{Q} z +(\frac{m_c}{n_c}r-
\frac{m_d}{n_d}s ), \frac{M_3'}{Q} \tau\right),
\nonumber \\
\psi^{t,M_4}_{ps} &=&
C_{ps}e^{-\pi \frac{M_4'}{Q}y^2_5}
\jtheta{ t'/M_4' \\ 0 }\left( \frac{M_4'}{Q} z +(\frac{m_a}{n_a}p-
\frac{m_d}{n_b}s ), \frac{M_4'}{Q} \tau\right),
\nonumber 
\end{eqnarray}
where $Q$ is defined as $Q={\rm l.c.m.}(n_a,n_b,n_c,n_d)$.
First, the product of $\psi^{j,M_1}_{pq}$ and $\psi^{k,M_2}_{qr} $
becomes
\begin{eqnarray}
\psi^{j,M_1}_{pq} \psi^{k,M_2}_{qr} 
=
C_{pq}C_{qr}
e^{-\pi \frac{M'}{Q}y^2_5}
\sum_{m\in Z_{M'}} \jtheta{ \frac{j'+k'+M_1'm}{M'} \\ 0 }
\left( \frac{M'}{Q} z +(\frac{m_a}{n_a}p-\frac{m_c}{n_c}r ), 
\frac{M'}{Q} \tau\right) \nonumber \\
\ \ 
\times
\jtheta{\frac{M_2'j'-M_1'k'+M_1'M_2'm}{M_1'M_2'M'}}
\left(
M_2'(\frac{m_a}{n_a} p-\frac{m_b}{n_b} q)-
M_1'(\frac{m_b}{n_b}q-\frac{m_c}{n_c}r), \frac{M_1' M_2' M'}{Q} \tau
\right),
\end{eqnarray}
where $M'=M_1'+M_2'$.
Then we repeat this product for $\psi^{l,M_3}_{rs}$ and 
use the orthogonal condition for the $M_4'$ sector 
because the relations $M_1'+M_2'+M_3'=M'+M_3'=M_4'$ hold by definition.
Finally we obtain the overlap integral for four
wave functions as
\begin{eqnarray}
&&
Y^{jklt}_{pqrs}
=
C^j_{pq}C^k_{qr}C^l_{rs}(C^t_{ps})^*
Q\sqrt{\frac{M_4'}{Q}}
\sum_{m\in Z_{M'}} \sum_{n\in Z_{M_4'}}
\delta_{j'+k'+M_1'm+l'+M'n, t'(\mod{M_4'})} \nonumber \\
&& 
\ \  \times
\jtheta{\frac{M_2'j'-M_1'k'+M_1'M_2'm}{M_1'M_2'M'}}
\left(
M_2'(\frac{m_a}{n_a} p-\frac{m_b}{n_b} q)-
M_1'(\frac{m_b}{n_b}q-\frac{m_c}{n_c}r), \frac{M_1' M_2' M'}{Q} \tau
\right)  \\
&& 
\ \ \times
\jtheta{\frac{M_3'(j'+k'+M_1'm)-M'l'+M'M_4'n}{M_3'M_4'M'}}
\left(
M_3'(\frac{m_a}{n_a} p-\frac{m_c}{n_c} r)-
M'(\frac{m_c}{n_c}r-\frac{m_d}{n_d} s), \frac{M' M_3' M_4'}{Q} \tau
\right) .      \nonumber
\end{eqnarray}
This result is just the product of two theta functions.
By solving the Kronecker delta, 
we obtain the sum of two theta functions like $\sum_m
y^{j'k'm}y^{l't'm'}$.
Therefore even including the non-Abelian Wilson lines we obtain 
results which are similar to Ref.~\cite{Abe:2009dr} 
for general four point couplings.

\end{document}